\documentclass[11pt,a4paper]{article}

\usepackage{amsmath, amssymb}
\usepackage[italicdiff]{physics}

\usepackage{graphicx}
\usepackage{comment}
\usepackage{jcappub}

\begin{document}

\title{Affleck-Dine inflation in supergravity} 

\author[a,b]{Masahiro Kawasaki,}
\author[a]{and Shusuke Ueda}

\affiliation[a]{ICRR, University of Tokyo, Kashiwa, 277-8582, Japan}
\affiliation[b]{Kavli IPMU (WPI), UTIAS, University of Tokyo, Kashiwa, 277-8583, Japan}

\emailAdd{kawasaki@icrr.u-tokyo.ac.jp}

\abstract{
Affleck-Dine inflation is a recently proposed model
in which a single complex scalar field, nonminimally coupled to gravity,
drives inflation
and simultaneously generates the baryon asymmetry of universe via Affleck-Dine mechanism.
In this paper we investigate
the supersymmetric implementation of Affleck-Dine inflation
in the use of two chiral superfields with appropriate superpotential and K\"ahler potential.
The scalar potential
has a similar form to the potential of original Affleck-Dine inflation,
and it gives successful inflation and baryogenesis.
We also consider the isocurvature perturbation
evolving after crossing the horizon,
and find that it is ignorable and hence consistent with the observations.
}


\maketitle

\section{Introduction}


Cosmological inflation is a hypothetical epoch of exponential expansion at the very beginning of the universe.
Inflation is driven by some slow-rolling scalar field, called inflaton.
There are many classes of theoretical models describing the origin of inflation, which will be tested by comparing with CMB observations~\cite{Akrami:2018odb}.
Among many inflation models, chaotic inflation with a power-law potential is a simple idea, but it cannot be consistent with observations since it generates too large tensor fluctuations ($=$ gravitational waves).
Several modified models to avoid this problem have been proposed.
One of such modifications is to add nonminimal coupling of the inflaton to gravity~\cite{Okada:2010jf,Linde:2011nh}, which makes the inflaton potential shallower and suppressed the tensor mode.

The origin of the baryon asymmetry of the universe is also an unsolved problem in particle physics and cosmology.
Since inflation dilutes preexisting baryon asymmetry, we need some mechanizm to produce baryon number after inflation.
Affleck-Dine (AD) mechanism~\cite{Affleck:1984fy} is one of promising baryogenesis models and realized in the minimal supersymmetric standard model (MSSM)~\cite{Dine:1995kz} (Refs.~\cite{Dine:2003ax,Allahverdi:2012ju} for review).
It is well known that MSSM has flat directions in the potential of the scalar fields (squarks, sleptons, Higgs).
Since the flat directions generally contain squarks, they have a baryon number.
The AD mechanism utilizes one of the flat directions (called AD field) and can produce the baryon asymmetry through dynamics of the AD field.

Recently Affleck-Dine inflation~\cite{Cline:2019fxx} was proposed.
In this model a complex scalar field with nonminimal coupling to the gravity drives inflation and simultaneously produces baryon asymmetry of the universe via AD mechanism
(For early attempts at explaining both inflation and baryogenesis by baryon number carrying scalar field, see Refs.~\cite{Hertzberg:2013mba,Takeda:2014eoa}, also Ref.~\cite{Lloyd-Stubbs:2020sed}). 
However, the model in Ref.~\cite{Cline:2019fxx} is based on non-supersymmetric (non-SUSY) potential and hence not so motivated as SUSY AD models.  
In this paper we construct an AD inflation model in the framework of SUSY (supergravity).\footnote{
Another model was proposed in~\cite{Lin:2020lmr}.}
We find that successful inflation and baryogenesis are realized by introducing appropriate superpotential and K\"ahler potential with nonminimal coupling to gravity.

The organization of this paper is as follows.
In Sec.~\ref{sec:ADinflation},  we explain the SUSY AD inflation model. 
The setup of the numerical calculation of dynamics of the AD field and its perturbations is described in Sec.~\ref{sec:analysis}, and the result is also shown in the section.
Finally, we conclude the paper in Sec.~\ref{sec:conclusion}.

\section{AD inflation}
\label{sec:ADinflation}

In this section we introduce the model of SUSY AD inflation and the materials required in the following analysis.
Most of the calculations follow the notation of Ref.~\cite{Cline:2019fxx}.

\subsection{Non-SUSY model}

Before presenting the SUSY AD inflation model we briefly reiew the non-SUSY AD inflation model in Ref.~\cite{Cline:2019fxx}.
In the Jordan frame the potential of the AD inflation is written as
\begin{align}
	V_J = m_\Phi^2|\Phi|^2 + \lambda|\Phi|^4 
	+ i\lambda'(\Phi^4 - \bar\Phi^4), 
	\label{eq.adinf-potential}
\end{align}
where $\Phi$ is the AD field, $\lambda$ and $\lambda'$ are the coupling constants.
For successful inflation we need to add a nonminimal matter-gravity coupling
\begin{align}
	\mathcal{L}_\mathrm{NM} = \frac{1}{2}R(1 + 2\xi|\Phi|^2),
\end{align}
where $\xi$ is the constant.
In this paper we use natural units, especially we set the reduced Planck mass $m_\mathrm{Pl} = 2.4 \times 10^{18}\ \mathrm{GeV}$ to unity.
By making a Weyl transformation, we can get the nearly flat potential in the Einstein frame
\begin{align}
	V_E = \frac{\lambda|\Phi|^4}{(1 + 2\xi|\Phi|^2)^2} + \cdots, 
	\label{eq.ad-einstein}
\end{align}
where the dots denote the rest of the potential and the noncanonical kinetic terms.
This form of potential suppresses the gravitational waves, and after the inflation, the CP violating terms in~\eqref{eq.adinf-potential} generate baryon number.


\subsection{SUSY Potential}

In the SUSY AD mechanism the potential of the AD field is lifted by the nonrenormalizable superpotential such as
\begin{align}
    W = \lambda\Phi^{p + 1} \quad (p \ge 3), \label{eq.ad-super}
\end{align}
and the soft SUSY-breaking mass term, and  the scalar potential is written as
\begin{align}
    V_J = m_\Phi^2|\Phi|^2 + (p + 1)^2|\lambda\Phi^p|^2 
	+ \pqty{A\lambda\Phi^{p + 1} + \mathrm{h.c.}}, \label{eq.ad-scalar}
\end{align}
where the last term is called $A$-term which breaks the baryon number.
Here and hereafter we use the same character $\Phi$ for a superfield and its scalar component.
Note that the form of the potential is similar to that of AD inflation~\eqref{eq.adinf-potential}.

If we can add nonminimal coupling
\begin{align}
    \mathcal{L}_\mathrm{NM} \sim R\Phi^{p},
\end{align}
we will get a nearly flat potential in Einstein frame, and after the inflation the A-term in the scalar potential~\eqref{eq.ad-scalar} will generate baryon asymmetry via usual AD mechanism.
The nonminimal coupling of scalar fields is realized in supergravity as studied in Refs.~\cite{Einhorn:2009bh, Lee:2010hj, Kallosh:2010ug, Kallosh:2010xz, Ketov:2012jt}.
	
	
However, it can be shown that the simple power-law superpotential~\eqref{eq.ad-super} accompanied by nonminimal coupling to supergravity cannot generate the nearly flat potential like~\eqref{eq.ad-einstein}.
Let us see this.
The nonminimal coupling in supergravity is realized as the specific form of K\"ahler potential~\cite{Ketov:2012jt}
\begin{align}
    K &= -3\ln\bqty{e^{-K_0/3} + \xi(\Phi^p + \bar\Phi^p)} \notag \\
      &= |\Phi|^2 - 3\ln\Omega^{-2}, \label{eq.kahler-nonminimal}
\end{align}
where $K_0 = |\Phi|^2$ is a minimal K\"ahler potential
(in the case without the nonminimal coupling) and
\begin{align}
    \Omega^2 \equiv \bqty{1 + \xi e^{K_0/3}(\Phi^p + \bar\Phi^p)}^{-1}
    \label{eq.Omega2}
\end{align}
represents the effect of the nonminimal coupling.
This replacement of $K_0$ to $K$ is equivalent to the Weyl transformation in non-SUSY theory.
The scalar potential in Einstein frame is calculated from the replaced K\"ahler potential~\eqref{eq.kahler-nonminimal} and the superpotential~\eqref{eq.ad-super} automatically.

To calculate the scalar potential,
we need the derivatives of the K\"ahler potential~\eqref{eq.kahler-nonminimal}:
\begin{align}
    K_\Phi &= \Omega^2\pqty{\bar\Phi - 3p\xi e^{K_0/3}\Phi^{p - 1}}, 
    \label{eq.KPhi-general} \\
    K_{\Phi\bar\Phi} &= \Omega^2 - \frac{1}{3}\Omega^4\xi e^{K_0/3}
    \bqty{
        (3p + |\Phi|^2)(\Phi^p + \bar\Phi^p) 
        - 9p^2\xi e^{K_0/3}|\Phi|^{2p - 2}
    }.
    \label{eq.KPhiPhi-general}
\end{align}
The scalar potential can be obtained as~\cite{Cremmer:1978hn}
\begin{align}
    V = e^K\pqty{\frac{|W_\Phi + K_\Phi W|^2}{K_{\Phi\bar\Phi}} - 3|W|^2}
\end{align}
where we ignore the D-terms
and the superpotential $W$ is given by Eq.~\eqref{eq.ad-super}.
Substituting~\eqref{eq.KPhi-general} and~\eqref{eq.KPhiPhi-general},
we can see there are many terms from the derivatives by $\Phi$
and we cannot obtain the simple potential like~\eqref{eq.ad-einstein}.

In this paper, to obtain the potential similar to~\eqref{eq.ad-einstein},
we introduce two chiral superfields $\Phi$ and $\Psi$ whose K\"ahler potential and the superpotential are given by
\begin{align}
    K_0  &= |\Phi|^2 + |\Psi|^2 ,
    \label{eq.kahler-minimal} \\
    W  &= \lambda\Phi^p\Psi + \mu\Phi^q.
    \label{eq.super}
\end{align}
We only consider the nonrenormalizable superpotentials with $p \ge 3$ and $q \ge 4$.
The couplings $\lambda$ and $\mu$ are typically of the order $1$, and for simplicity we assume they are real
(Notice that in our unit the reduced Planck mass $m_\mathrm{Pl}$ is $1$).
Throughout this paper, we assume that the superfields $\Psi$ is stabilized at $\Psi = 0$ during and after the inflation.
Thanks to this stabilization, as we will see later, the first term of the superpotential $\lambda\Phi^p\Psi$ provides a nearly flat potential for $\Phi$, while the second term $\mu\Phi^q$ contributes to A-terms which is necessary to generate the baryon asymmetry.
	
As we mentioned above,
we can introduce the nonminimal coupling to supergravity
by replacing the K\"ahler potential~\eqref{eq.kahler-minimal}
\begin{align}
    K_0 \to K = K_0 - 3\ln\Omega^{-2},
    \label{eq.kahler}
\end{align}
where
\begin{align}
    \Omega^2 = \bqty{1 + \xi e^{K_0/3}(\Phi^p + \bar\Phi^p)}^{-1}
\end{align}
is the same as~\eqref{eq.Omega2} except $K_0$.

In the following analysis, we need the partial derivatives of the K\"ahler potential~\eqref{eq.kahler} by the chiral superfields $\Phi$ and $\Psi$.
The first derivative $K_\Phi$ is given as~\eqref{eq.KPhi-general} and
\begin{align}
    K_\Psi &= \Omega^2\bar\Psi.
\end{align}
The second derivatives are Eq.~\eqref{eq.KPhiPhi-general} and
\begin{align}
    K_{\Psi\bar\Psi} &= \Omega^2 - \frac{1}{3}\Omega^4\xi e^{K_0/3}
        |\Psi|^2(\Phi^p + \bar\Phi^p),
    \label{eq.KPsiPsi-general} \\
    K_{\Phi\bar\Psi} &= -\frac{1}{3}\Omega^4\xi e^{K_0/3}\Psi
        \bqty{
			\bar\Phi(\Phi^p + \bar\Phi^p) + 3p\Phi^{p - 1}
		}.
	\label{eq.KPhiPsi-general}
\end{align}
Because $\Psi$ is stabilized at $\Psi = 0$, the second derivatives~\eqref{eq.KPhiPhi-general}, \eqref{eq.KPsiPsi-general} and~\eqref{eq.KPhiPsi-general} can be evaluated as
\begin{align}
    K_{\Phi\bar\Phi} &= \Omega^2 - \frac{1}{3}\Omega^4\xi e^{|\Phi|^2/3}
	    \bqty{
			(3p + |\Phi|^2)(\Phi^p + \bar\Phi^p) - 9p^2\xi e^{|\Phi|^2/3}|\Phi|^{2p - 2}
		}, \\
	K_{\Psi\bar\Psi} &= \Omega^2, 
	\label{eq.KPsiPsi} \\
	K_{\Phi\bar\Psi} &= 0 
	\label{eq.KPhiPsi},
\end{align}
where
\begin{align}
	\Omega^2 = \bqty{1 + \xi e^{|\Phi|^2/3}(\Phi^p + \bar\Phi^p)}^{-1}.
\end{align}
	
Generally, with multiple superfields, the scalar potential of supergravity is obtained by~\cite{Cremmer:1978hn}
\begin{align}
	V = e^K\pqty{D_I W K^{I\bar J}\overline{D_J W} - 3|W|^2},
\end{align}
where we again ignore the D-terms.
The subscripts $I$ and $J$ represent the partial derivatives by $\Phi$ or $\Psi$.
The covariant derivatives $D_I W$ stand for $W_I + K_I W$, and $K^{I\bar J}$ is the inverse matrix of $K_{I\bar J}$.
When $\mu = 0$ we get $W = 0$ and $D_\Phi W = 0$, and the scalar potential is given by
\begin{align}
	V = e^K K^{\Psi\bar\Psi}|D_\Psi W|^2.
	\label{eq.scalar-potential}
\end{align}
Note that the matrix $K_{I\bar J}$ is diagonal	(See Eq.~\eqref{eq.KPhiPsi}).
Using Eq.~\eqref{eq.kahler}, \eqref{eq.KPsiPsi} and $D_\Psi W = \lambda\Phi^p$,
the scalar potential is rewritten as
\begin{align}
	V \simeq 
	    \frac{\lambda^2|\Phi|^{2p}}{\bqty{1 + \xi (\Phi^p + \bar\Phi^p)}^2} 
	    + m_\Phi^2|\Phi|^2.
	\label{eq.potential}
\end{align}
Here we add the SUSY breaking soft mass term in the potential.
When $\mu \ne 0$ the superpotential $W \ne 0$ during and after the inflation, which leads to the A-terms in the potential.
Finally we obtain 
\begin{align}
	V &\simeq \frac{\lambda^2|\Phi|^{2p}}{\bqty{
		1 + \xi (\Phi^p + \bar\Phi^p)}^2} + m_\Phi^2|\Phi|^2 + \mu(A\Phi^q + \mathrm{h.c.}).
	\label{eq.potential-a}
\end{align}
The SUSY breaking scales $m_\Phi$ and $A$ are typically the order of $O(1-100)\ \mathrm{TeV}$ [$O(10^{-15} - 10^{-13})$ in our units].
Here and hereafter we require that the field value should be smaller than the Planck mass to avoid supergravity effect\footnote{
In the region with the large field value $|\Phi| \gtrsim 1$ the inflaton field rapidly falls due to the exponential factor in the potential~\eqref{eq.scalar-potential} and we cannot obtain sufficient e-folds.}
and hence we ignore $|\Phi|^2$ compared to $1$.
When $\mu \ne 0$ many other terms appear in the potential~\eqref{eq.potential-a}, but we can also ignore them since their contribution to the dynamics are not important, which we can check by numerical analysis using full potentials.
	
\subsection{Equations of motion}

Let us consider evolution of the spatially homogeneous fields.
The Lagrangian for the inflaton field $\Phi$ is
\begin{align}
	\mathcal{L} = -K_{\Phi\bar\Phi}\partial_\mu\Phi\partial^\mu\bar\Phi - V
	= \frac{1}{2}z^2(\dot\phi^2 + \dot\chi^2) - V,
\end{align}
where we write the real and imaginary parts of $\Phi$ as $\Phi =(\phi + i\chi)/\sqrt{2}$ and set $z^2 = K_{\Phi\bar\Phi}$.
	
To obtain the time evolution of the system, we introduce the canonical fields $u$ and $v$ by
\begin{align}
	\pmqty{\dot u \\ \dot v}
		= z\pmqty{\cos\psi & \sin\psi \\ -\sin\psi & \cos\psi}\pmqty{\dot\phi \\ \dot\chi}.
	\label{eq.uv-definition}
\end{align}
The rotation by the angle $\psi$ is not necessary just to obtain the canonical fields, but it is important to correctly evaluate the second derivatives of the potential $V$~\cite{Cline:2019fxx}.
Note that the second derivatives of the new fields $u$ and $v$ by the original fields $\phi$ and $\chi$ are given by $u_{\phi\chi} = (z\cos\psi)_\chi = z_\chi\cos\psi - z\psi_\chi\sin\psi$ and so on, but they do not satisfy the relations $u_{\phi\chi} = u_{\chi\phi}$ or $v_{\phi\chi} = v_{\chi\phi}$ generally.
On the other hand, if we set the derivatives of the rotation angle $\psi$ as
\begin{align}
	\pmqty{\psi_\phi \\ \psi_\chi}
	= \frac{1}{z}\pmqty{z_\chi \\ -z_\phi},
	\label{eq.psii}
\end{align}
the second derivatives always commute, i.e. the transformation matrix in Eq.~\eqref{eq.uv-definition}
\begin{align}
	Z = z\pmqty{\cos\psi & \sin\psi \\ -\sin\psi & \cos\psi}
\end{align}
can be considered as the Jacobian.
Using $\psi$, the derivatives of the potential~\eqref{eq.potential} or~\eqref{eq.potential-a} by the canonical fields $u$ and $v$ can be calculated as
\begin{align}
		V_k &= V_i{Z^{-1}}_{ik},\label{eq.Vk} \\
		V_{kl} &= V_{ij}{Z^{-1}}_{ik}{Z^{-1}}_{jl} + V_i({Z^{-1}}_{ik})_j{Z^{-1}}_{jl} \label{eq.Vkl},
\end{align}
where the subscripts $i$ and $j$ stand for the derivatives by the original fields $\phi$ and $\chi$, while $k$ and $l$ for $u$ and $v$.
		
The equations of motion of the canonical fields $u$ and $v$ can be obtained as
\begin{gather}
	\ddot u + 3H\dot u + V_u = 0, \label{eq.eom-u} \\
	\ddot v + 3H\dot v + V_v = 0,\label{eq.eom-v}
\end{gather}
where $H$ is the Hubble parameter and can be evaluated using the energy density of the universe $\rho$ as
\begin{align}
	H^2 = \frac{1}{3}\rho 
	    = \frac{1}{3}\bqty{\frac{1}{2}(\dot u^2 + \dot v^2) +V}.
	\label{eq.hubble}
\end{align}
The time evolution of the rotation angle $\psi$ is obtained by Eq.~\eqref{eq.psii} as
\begin{align}
	\dot\psi = -\frac{z_\phi\dot\chi - z_\chi\dot\phi}{z}.
	\label{eq.eom-psi}
\end{align}
Eqs.~\eqref{eq.eom-u}, ~\eqref{eq.eom-v} and~\eqref{eq.eom-psi} are the fundamental equations of motion of our system.

\subsection{Perturbations}

To calculate the perturbations and the important variables such as the spectral index or the tensor-to-scalar ratio, we have to rorate $u$ and $v$ again into adiabatic and entropy directions as~\cite{Gordon:2000hv,Byrnes:2006fr}
\begin{align}
	\pmqty{\dot\sigma \\ \dot s}
		= \pmqty{\cos\theta & \sin\theta \\ -\sin\theta & \cos\theta}
			\pmqty{\dot u \\ \dot v},
	\label{eq.sigmas-definition}
\end{align}
where the rotation angle $\theta$ is given by
\begin{align}
	\theta = \tan^{-1}\frac{\dot v}{\dot u}.
\end{align}
We can see the adiabatic direction $\dot\sigma$ is parallel to the inflationary trajectory, while the entropy direction $\dot{s}$ is orthogonal to this.
	
Using the new fields $\sigma$ and $s$,	the slow-roll parameters are defined as 
\begin{align}
	\epsilon_m &= \frac{1}{2}\pqty{\frac{V_m}{V}}^2, \\
	\eta_{mn} &= \frac{V_{mn}}{V}.
\end{align}
Here the subscripts $m$ and $n$ in RHS shows the derivatives by $\sigma$ or $s$, which can be evaluated using the derivatives by the original canonical fields $u$ and $v$ (Eq.~\eqref{eq.Vk} and~\eqref{eq.Vkl}) as~\cite{Gordon:2000hv} 
\begin{align}
	V_\sigma &= V_u\cos\theta + V_v\sin\theta, \\
	V_s &\simeq 0, \\
	V_{\sigma\sigma} &= V_{uu}\cos^2\theta
		+ 2V_{uv}\cos\theta\sin\theta + V_{vv}\sin^2\theta, \\
	V_{ss} &= V_{uu}\sin^2\theta
		- 2V_{uv}\sin\theta\cos\theta + V_{vv}\cos^2\theta, \\
	V_{\sigma s} &= -(V_{uu} - V_{vv})\cos\theta\sin\theta
		+ V_{uv}(\cos^2\theta - \sin^2\theta).
\end{align}

The spectral index $n_s$ of scalar perturbation and the tensor-to-scalar ratio $r$ are~\cite{Byrnes:2006fr}
\begin{align}
	n_s &= 1 - 6\epsilon_\sigma\pqty{1 - \frac{2}{3}\sin^2\Delta}
	    + 2\eta_{\sigma\sigma}\cos^2\Delta
		+ 4\eta_{\sigma s}\cos\Delta\sin\Delta + 2\eta_{ss}\sin^2\Delta, \\
	r &= 16\epsilon_\sigma,	
\end{align}
where
\begin{align}
	\sin\Delta &= -2c\eta_{\sigma s},
	\quad c = 2 - \ln 2 - \gamma = 0.73, \\
	\cos\Delta &= +\sqrt{1 - \sin^2\Delta}
\end{align}
($\gamma = 0.57$ is Euler-Mascheroni constant).
The power spectrum of the scalar perturbation is
\begin{align}
	P_\mathcal{R} = \frac{V}{24\pi^2\epsilon_\sigma}
	    \bqty{
		    1 - 2\epsilon_\sigma
		    + 2c(3\epsilon_\sigma - \eta_{\sigma\sigma} 
		    - 2\eta_{\sigma s}T_\mathcal{RS})
		},
	\label{eq.PR}
\end{align}
where $T_\mathcal{RS}$ is a off-diagonal component of the transfer function, which represents the time evolution of the perturbation after exiting the horizon.
The definition of $T_\mathcal{RS}$ will be given later.
For successful inflation consistent with observations, $n_s = 0.965 \pm 0.004$, $r < 0.1$ and $P_\mathcal{R} = 2.1 \times 10^{-9}$ are required~\cite{Akrami:2018odb}.
	
The Fourier modes of perturbations $\delta u$ and $\delta v$ of the canonical fields $u$ and $v$ in the spatially flat gauge satisfy the following equations of motion~\cite{Gordon:2000hv}:
\begin{align}
	\ddot{\delta u}_k &+ 3H\dot{\delta u}_k + \frac{k^2}{a^2}\delta u_k 
	+ \bqty{
		V_{kl} - \frac{1}{a^3}\dv{t}\pqty{
			\frac{a^3}{H}\dot u_k\dot u_l
		}
	}\delta u_l = 0
	\qquad (u_k, u_l = u, v),
\end{align}
where $a$ is the scale factor of the universe.
Since we are interested in the evolution after exiting the horizon, we ignore the third term of LHS of the above equation.
By carrying out the time derivative in the bracket, we obtain 
\begin{align}
	\ddot{\delta u}_k &+ 3H\dot{\delta u}_k + \bqty{
		V_{kl} - 3\dot u_k\dot u_l
		- \frac{1}{2H^2}\dot u_k\dot u_l(\dot u^2 + \dot v^2)
		- \frac{1}{H}(\ddot u_k\dot u_l + \dot u_k\ddot u_l)
	}\delta u_l = 0.
	\label{eq.perturbation-eom}
\end{align}
Here we have used 
\begin{align}
	\dot H = -\frac{1}{2}(\dot u^2 + \dot v^2),
\end{align}
which is obtained from  Eqs.~\eqref{eq.hubble}, \eqref{eq.eom-u} and \eqref{eq.eom-v}.

We can obtain the adiabatic/entropy perturbations by the same rotation as the background field~\eqref{eq.sigmas-definition},
\begin{align}
	\pmqty{\delta\sigma \\ \delta s}
	= \pmqty{\cos\theta & \sin\theta \\ -\sin\theta & \cos\theta}
		\pmqty{\delta u \\ \delta v}.
\end{align}
It is convenient to use the dimensionless curvature perturbation $\mathcal{R}$ and the isocurvature perturbation $\mathcal{S}$ defined by
\begin{align}
	\mathcal{R} = \frac{H}{\dot\sigma}\delta\sigma,
	\quad \mathcal{S} = \frac{H}{\dot\sigma}\delta s,
\end{align}
where $\dot\sigma = \sqrt{\dot u^2 + \dot v^2}$.
The transfer functions are defined as the correlation functions between the curvature/isocurvature perturbations when crossing the horizon and them at the some time we consider after exiting the horizon~\cite{Cline:2019fxx, Byrnes:2006fr}:
\begin{align}
	\pmqty{\mathcal{R} \\ \mathcal{S}}
	= \pmqty{T_\mathcal{RR} & T_\mathcal{RS} \\ T_\mathcal{SR} & T_\mathcal{SS}}
		\pmqty{\mathcal{R}_* \\ \mathcal{S}_*}.
	\label{eq.transfer}
\end{align}
where the subscript $_*$ shows they are evaluated at the horizon crossing.
In the slow-roll limit, $T_\mathcal{RR} = 1$ since the curvature perturbation is conserved after exiting the horizon, and $T_\mathcal{SR} = 0$ because the curvature perturbation does not generate the isocurvature one.
The autocorrelation of the isocurvature perturbation $T_\mathcal{SS}$ is generally ignorable.
The component $T_\mathcal{RS}$ is important and we need $|T_\mathcal{RS}| \lesssim 0.1$ to satisfy the constraint of observations~\cite{Akrami:2018odb}.

\subsection{Baryon asymmetry}

Suppose that the superfield $\Phi$ carry non-zero baryon number $B$.
The baryon number density is given as usual by
\begin{align}
	n_B = -iB(\Phi\dot{\bar\Phi} - \bar\Phi\dot\Phi) = 2B\Im\Phi\dot{\bar\Phi}.
\end{align}
Let us introduce the ratio of this to the number density of $\Phi$, $n_\Phi = \rho/m_\Phi$, as
\begin{align}
	\eta \equiv \frac{n_B}{n_\Phi}.
	\label{eq.baryon-Phi}
\end{align}
$\eta$ is expected to become a constant $\eta_c$ after inflation and until the time of reheating
\footnote{Ref.~\cite{Lozanov:2014zfa} points out the effect of nonlinear dynamics at the end of inflation (preheating) in this kind of models.
We do not consider this effect here.}
.
If so, the baryon-to-entropy ratio at the time of reheating (RH) can be evaluated as
\begin{align}
	\eta_B^\mathrm{RH}
		= \frac{n_B^\mathrm{RH}}{s_\mathrm{RH}}
		= \eta_c\frac{n_\Phi^\mathrm{RH}}{s_\mathrm{RH}},
	\label{eq.baryon-entropy-general}
\end{align}
where $s$ is the entropy density given by $s_\mathrm{RH} = (2\pi^2/45)g_* T_\mathrm{RH}^3$ with $g_*$ the effective degree of freedom and $T_\mathrm{RH}$ the reheating temperature.
Note that the energy density of the universe at reheating $\rho_\mathrm{RH}$ and the decay rate of the field $\Gamma_\Phi$ satisfy the following relations~\cite{Kofman:1997yn}:
\begin{align}
	\rho_\mathrm{RH} = \frac{4}{3}\Gamma_\Phi^2,
	\quad \Gamma_\Phi^2 = \frac{\pi^2 g_*}{90}T_\mathrm{RH}^4.
	\label{eq.GammaPhi-relations}
\end{align}

The observed baryon-to-entropy ratio can be evaluated as
\begin{align}
	\eta_B
		= \frac{1}{3}\alpha\eta_c\frac{T_\mathrm{RH}}{m_\Phi}
		\sim 0.1\eta_c\frac{T_\mathrm{RH}}{m_\Phi},
	\label{eq.baryon-entropy}
\end{align}
where $\alpha$ is a factor taking into account the effect of the sphaleron process~\cite{Harvey:1990qw} which converts some of the baryon number into lepton number.
To avoid the gravitino problems, the reheating temperature cannot be higher than about $10^8\ \mathrm{GeV}$~\cite{Kawasaki:2006gs}.
The observed baryon-to-entropy ratio is $\eta_B^\mathrm{obs} = 8.6 \times 10^{-11}$~\cite{Aghanim:2018eyx}.

\section{Numerical analysis}
\label{sec:analysis}

\subsection{Methods}

Based on the formula presented in the previous section we numerically calculate the evolution of the homogeneous $\Phi$ and its perturbations. 
As for the initial value of the $\Phi$, the scalar potential~\eqref{eq.potential} or~\eqref{eq.potential-a} must be positive for successful inflation, and the absolute value $|\Phi|$ be smaller than Planck mass ($1$ in our unit) to avoid supergravity effect.
In this paper we set the initial value of the time derivative $\dot\Phi$ to $0$.
We also set the initial value of the rotation angle $\psi$ to $0$.
	
We use the e-folding number $N$ as the time coordinate.
In this paper we set $N = 0$ at the beginning of inflation, and $N$ increases along with the time until the end of inflation $N = N_\mathrm{end} > 0$.
 We cannot determine the exact moment when inflation ends, but here we use the time the sign of $\phi$ or $\chi$ is flipped for the first time.
The energy density of that time is $\rho_\mathrm{end}$.
	
To compare with the observation, we need to determine when the CMB scale $k_* = 0.05\ \mathrm{Mpc^{-1}}$ or $0.002\ \mathrm{Mpc^{-1}}$ exits the horizon.
According to $N_\mathrm{end}$ and $\rho_\mathrm{end}$, the e-folding number of that time $N_*$ can be evaluated by the following equation~\cite{Akrami:2018odb, Liddle:2003as}:
\begin{align}
	N_\mathrm{end} - N_*
	= 67 - \ln\frac{k_*}{H_0}
	+ \frac{1}{4}\ln\frac{V_*^2}{\rho_\mathrm{end}}
	+ \frac{1}{12}\ln\frac{\rho_\mathrm{RH}}{g_*\rho_\mathrm{end}},	
\end{align}
where $H_0 = 67.4\ \mathrm{km\ s^{-1}\ Mpc^{-1}}$ is the present value of Hubble parameter and $V_*$ is the value of the potential $V$ when the scale $k_*$ exits the horizon.
Using Eq.~\eqref{eq.GammaPhi-relations} and substitute the value $k_* = 0.05\ \mathrm{Mpc^{-1}}$,
\begin{align}
	N_*= N_\mathrm{end} - 61.43 + \frac{1}{3}\ln\rho_\mathrm{end}
		- \frac{1}{3}\ln T_\mathrm{RH} - \frac{1}{2}\ln V_*.
\end{align}
If we use the scale $k_* = 0.002\ \mathrm{Mpc^{-1}}$, the constant of RHS will be $64.65$ instead of $61.43$.

Given $N_*$, we can evaluate the observables $n_s$, $r$ and $P_\mathcal{R}$.
The transfer function is needed to estimate the accurate value of the power spectrum, but in our situations the absolute value of $T_\mathcal{RS}$ is always small, so we ignore it here.
This assumption will be checked later.
	
Since the first term of the potential~\eqref{eq.potential-a} is leading during inflation, the potential is proportional to $\lambda^2$.
Although the dynamics of the background fields or the values of $n_s$ and $r$ do not depend on the coupling $\lambda$ so strong, the power spectrum depends proportionally $P_\mathcal{R} \propto \lambda^2$.
Thus, we can get the observed value $P_\mathcal{R}^\mathrm{obs} = 2.1 \times 10^{-9}$ by adjusting the coupling $\lambda$:
First we calculate $P_\mathcal{R}$ for an arbitrary $\lambda$, then replace it as
\begin{align}
	\lambda \to \pqty{\frac{P_\mathcal{R}^\mathrm{obs}}{P_\mathcal{R}}}^{1/2}\lambda.
	\label{eq.lambda-rec}
\end{align}
Because this replacing can affect the dynamics or the other observables a little, we need to repeat it for several times to obtain the precise estimation.
	
Once we find a parameter set for successful inflation, we proceed to evaluate the baryon-to-entropy ratio~\eqref{eq.baryon-entropy} and compare the value to the observed one.
We also need to check the the assumption we made above that the value of $T_\mathcal{RS}$ is ignorable.
The transfer function can be calculated by Eq.~\eqref{eq.perturbation-eom} and~\eqref{eq.transfer}.
	
We tried with the three sets of parameters showed in Table~\ref{table.parameters}.	In every Model we chose the relatively small initial value for the imaginary part $\chi_0$ compared to the real part $\phi_0$, since the potential~\eqref{eq.potential} or~\eqref{eq.potential-a} has a nearly flat direction along with the real axis $\chi = 0$.
We also let the SUSY breaking parameter $A$ have a complex phase in Model 2 and 3.
It is because, if without the phase, the potential is completely symmetric to the real axis and no baryon asymmetry will be generated.\footnote{
We can make $A$ real by choosing appropriate initial values $\phi_0$ and $\chi_0$ along with the other flat directions of the potential.
For example there are two other flat directions when $p = 3$, which is the case we have studied in this paper.}
\begin{table*}[h]
	\centering
	\caption{Parameters used in our analysis. 
	Since $\mu = 0$ in Model 1, $q$ and $A$ are not relevant and we did not show them.
	For each Model we started with $\lambda = 1$ and adjusted it using Eq.~\eqref{eq.lambda-rec} to get the observed value of the power spectrum.
	Notice that we set $m_\text{PL}$ to $1$, so $m_\Phi = 10^{-13} = 240$~TeV.}
	\label{table.parameters}
	\vspace{4mm}
	\begin{tabular}{ccccccccccc}
		\hline
		Model & $p$ & $q$ & $\mu$ & $\xi$
			& $m_\Phi$                            & $A$
			& $\phi_0$ & $\chi_0$ & $B$ & $T_\mathrm{RH}$ \\
		\hline\hline
		1   & $3$ & -   & $0$   & $10^5$
			& $10^{-13}$ & -
			& $0.12$   & $0.008$  & $1$ & $10^{-10}$ \\
		2   & $3$ & $4$ & $1$   & $10^5$
			& $10^{-13}$                          & $(1 + 0.2i) \times 10^{-13}$
			& $0.12$   & $0.008$  & $1$ & $10^{-10}$ \\
		3   & $3$ & $5$ & $1$   & $10^5$
			& $10^{-13}$                          & $(1 + 0.2i) \times 10^{-13}$
			& $0.12$   & $0.008$  & $1$ & $10^{-10}$ \\
		\hline
	\end{tabular}
\end{table*}

\subsection{Results}
\label{sec:result}

In every Model, after replacing $\lambda$ by Eq.~\eqref{eq.lambda-rec} three times, we get the observed value $P_\mathcal{R} = 2.1 \times 10^{-9}$ with the coupling $\lambda = 2.36$.
The end of inflation is at $N_\mathrm{end} = 90.9$, while the horizon crossing of the scale $k_* = 0.05\ \mathrm{Mpc^{-1}}$ is at $N_* = 40.1$.
The spectral index of scalar perturbation is $n_s = 0.961$ and the tensor-to-scalar ratio $r_{0.002} = 3.9 \times 10^{-3}$,	which are consistent with observations.\footnote{
We evaluated $n_s$ at $k_* = 0.05\ \mathrm{Mpc^{-1}}$, while $r$ at $k_* = 0.002\ \mathrm{Mpc^{-1}}$.}
Notice that these values depend on the reheating temperature $T_\mathrm{RH}$,
which we set to $10^{-10}$ ($\sim 10^8\ \mathrm{GeV}$) in every Model.
	
\begin{figure}[t]
	\centering
	\includegraphics[width=0.7\textwidth]{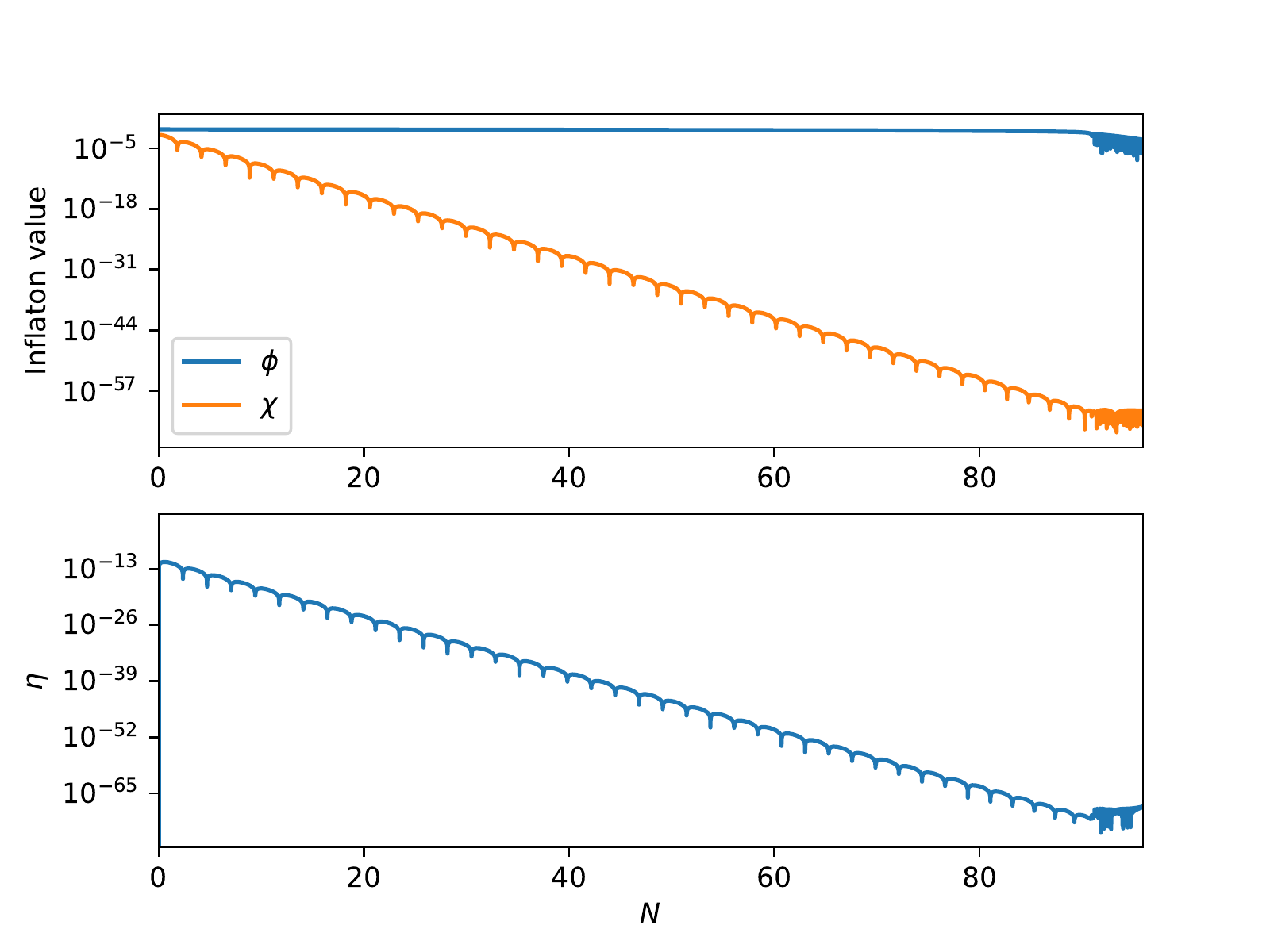}
	\caption{The time evolution of the background field $\phi$, $\chi$ and the baryon-to-$\Phi$ ratio for the parameters of Model 1 in Table~\ref{table.parameters}.}
	\label{fig.result-bg-1}
\end{figure}

\begin{figure}[ht]
	\centering
	\includegraphics[width=0.7\textwidth]{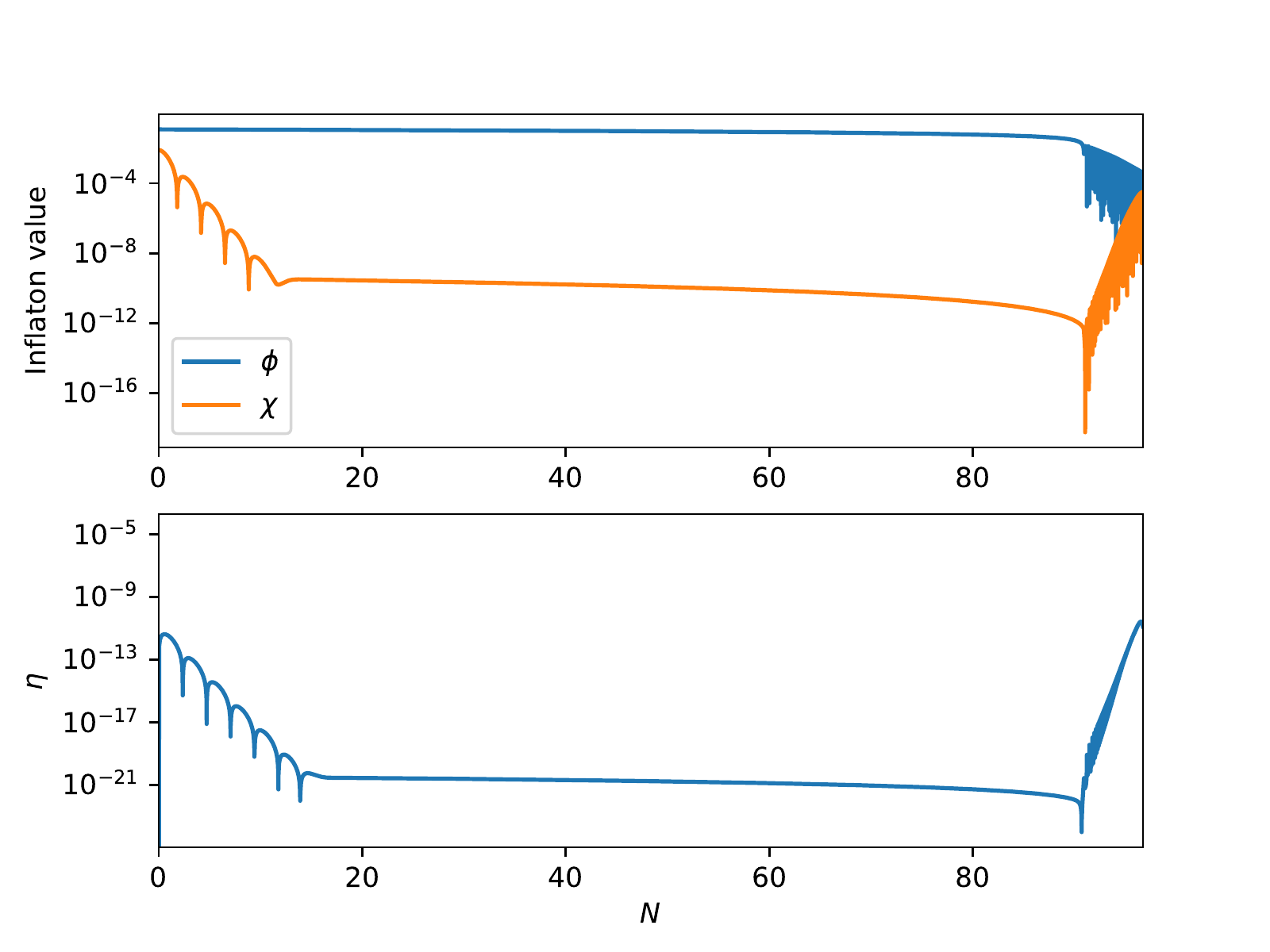}
	\caption{The time evolution of the background field $\phi$, $\chi$ and the baryon-to-$\Phi$ ratio for the parameters of Model 2 in Table~\ref{table.parameters}.}
	\label{fig.result-bg-2}
\end{figure}

\begin{figure}[ht]
	\centering
	\includegraphics[width=0.7\textwidth]{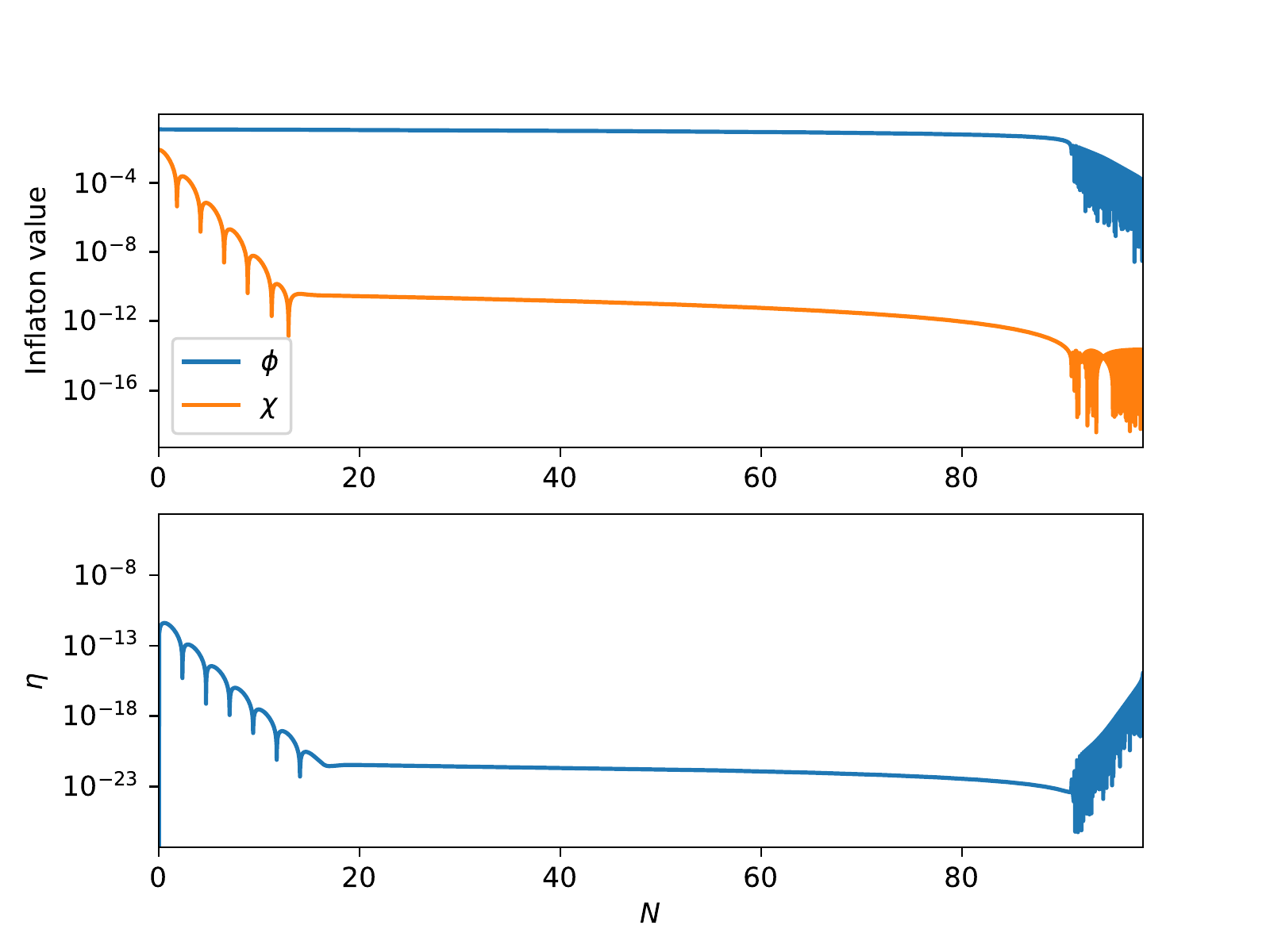}
	\caption{The time evolution of the background field $\phi$, $\chi$ and the baryon-to-$\Phi$ ratio for the parameters of Model 3 in Table~\ref{table.parameters}.
	We have not showed the evolution of $N \gtrsim 100$ because it is difficult to numerically calculate it due to the rapid oscillation of the fields.}
	\label{fig.result-bg-3}
\end{figure}

The time evolution of the original background fields $\phi$ and $\chi$ and the baryon-to-$\Phi$ ratio $\eta$ are shown in Fig.~\ref{fig.result-bg-1}, \ref{fig.result-bg-2} and~\ref{fig.result-bg-3}, respectively.
In every Model we can see the imaginary part $\chi$ of the complex scalar field falls to $0$ rapidly, and the real part $\phi$ plays the essential roll of inflaton.

In Model 1, without A-term, 
the scalar potential~\eqref{eq.potential} is completely symmetric to the real axis (= inflationary trajectory)
and the entropy direction $\chi$ oscillates with shrinking amplitude.
After inflation the AD field just oscillates along with the real axis
and no baryon asymmetry is generated.

On the other hand, in Model 2 and 3,
the oscillation of the imaginary part $\chi$ stops at some point during inflation
due to the complex phase of the SUSY breaking parameter $A$.
This phase make the AD field $\Phi$ evolve to the imaginary direction after inflation
and then the trajectory becomes circular.
With the parameters of Model 2,
we get $\eta_c \gtrsim 10^{-12}$ after inflation.
Using Eq.~\eqref{eq.baryon-entropy} and $T_\mathrm{RH} / m_\Phi = 10^3$, we can get $\eta_B \gtrsim 10^{-10}$, which is sufficient to explain the observed baryon asymmetry.	
It is difficult to make an accurate numerical calculation for Model 3, due to the rapid oscillation after inflation, but we can expect $\eta_c \gtrsim 10^{-16}$.

\begin{figure}[ht]
	\centering
	\includegraphics[width=0.45\textwidth]{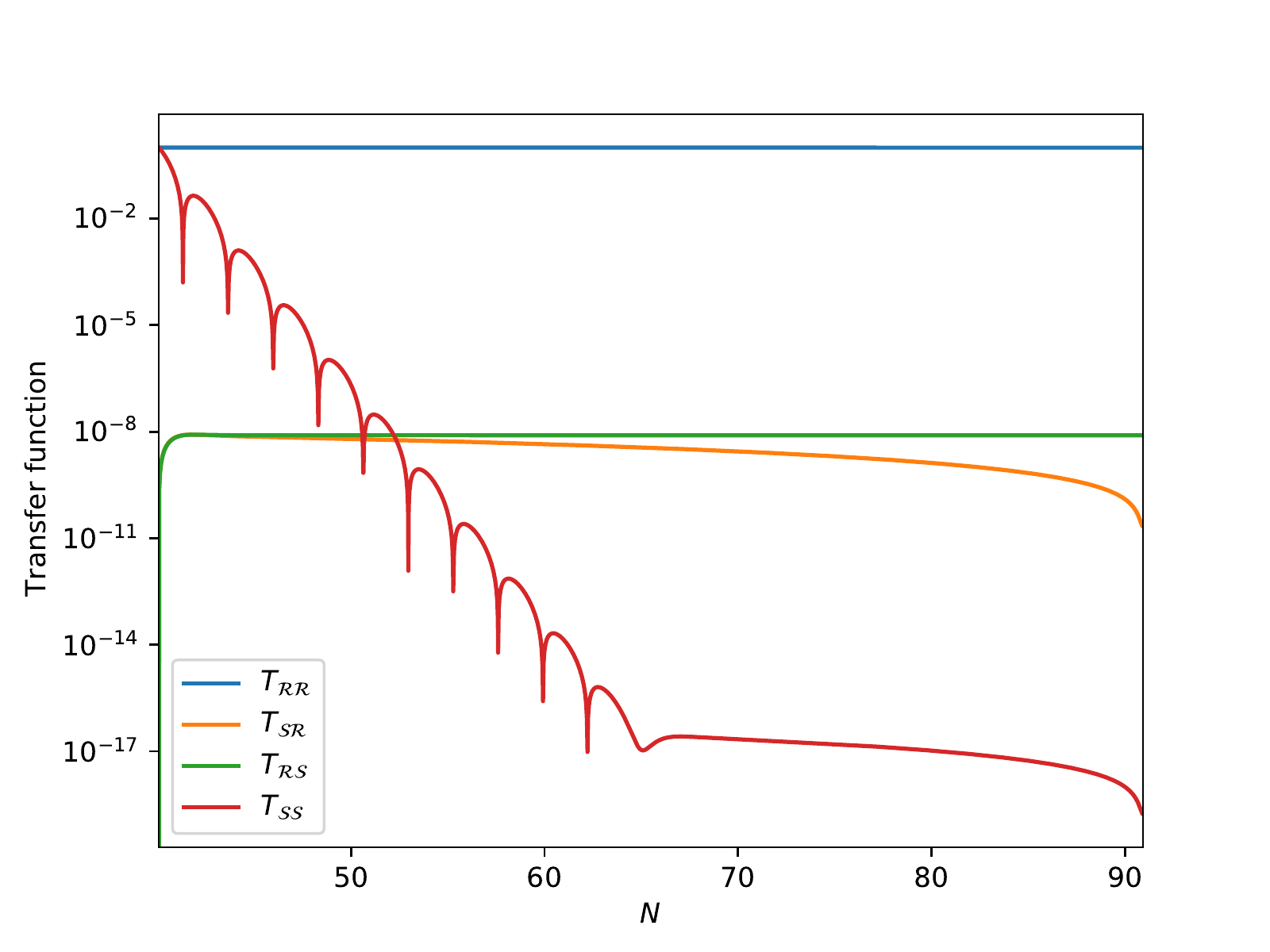}
	\includegraphics[width=0.45\textwidth]{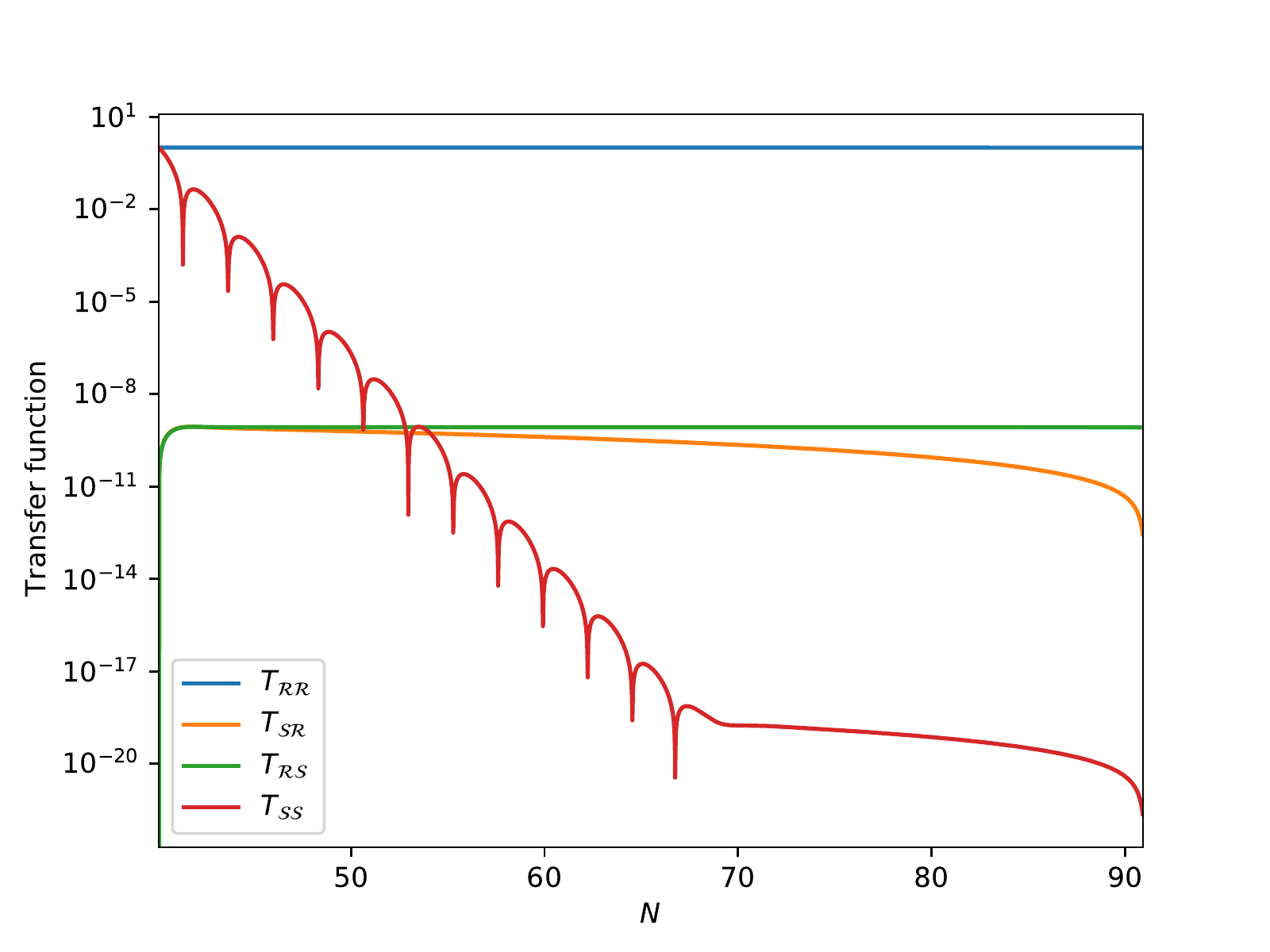}
	\caption{The transfer function of the adiabatic/entropy perturbation~\eqref{eq.transfer} after crossing the horizon for Model 2 (left) and Model~3 (right).
	We have not showed the evolution after the end of inflation
	because our analysis based on Ref.~\cite{Gordon:2000hv}
	does not include any higher-order corrections such as backreaction from small-scale perturbations.}
	\label{fig.result-pert-2}
\end{figure}

The transfer functions~\eqref{eq.transfer} of Model 2 and 3 are shown in Fig.~\ref{fig.result-pert-2}.
In both cases the adiabatic perturbation conserves ($T_\mathcal{RR} \simeq 1$)
and the cross correlation $T_\mathcal{SR}$ are small,
which are in accordance with the slow-roll prediction~\cite{Byrnes:2006fr}
$T_\mathcal{RR} = 1$ and $T_\mathcal{SR} = 0$.
The entropy autocorrelation $T_\mathcal{SS}$ rapidly falls
and it is difficult to directly observe it.
We can see $T_\mathcal{RS}$ is also very small, which justifies the assumption we made above that we can ignore this term when evaluating $P_\mathcal{R}$ with Eq.~\eqref{eq.PR}.
	
\section{Conclusions}
\label{sec:conclusion}
	
There have been several studies about whether a single complex scalar field plays a roll of inflaton and simultaneously generate the observed baryon asymmetry via Affleck-Dine mechanism.
Affleck-Dine inflation~\cite{Cline:2019fxx} is one of the successful scenarios.
We have construct an AD inflation model in the framework of supergravity with  use of nonminimal matter-supergravity coupling.
In the model we introduced two chiral superfields and the specific form of the superpotential and K\"ahler potential.
It was found that there are some regions in the parameter space where one of the scalar fields, $\Phi$, drives successful inflation and, after the inflation, the field rotates about the origin, generating the observed baryon asymmetry.
In our models the tensor-to-scalar ratio $r$ and the transfer function $T_\mathcal{RS}$, which describes the rate of transition from isocurvature perturbation to curvature perturbation, are both taking small values.
It is difficult to detect these signals within the sensitivity of the current experiments, but there may exist some other parts of the parameter space where $r$ or $T_\mathcal{RS}$ can have larger values.
Also the upcoming CMB experiments such as LiteBIRD~\cite{Hazumi:2019lys} can detect the presented value $r = 0.004$ in this paper.
	
We did not specify the origin of the superpotential and the mechanism that stabilizes the second superfield $\Psi$ at $0$, which will be studied in future works.
The large values of the nonminimal coupling $\xi$ we used also need to be explained.
	
\bibliography{main}

\end{document}